\documentclass[showpacs,prl,aps,twocolumn,float,floatfix]{revtex4}
\usepackage{graphicx,epsfig}
\usepackage{amsmath,amssymb}

\begin{document}

\title{Fractal Weyl laws for quantum decay in generic dynamical systems}
\author{Marten Kopp}
\author{Henning Schomerus}
\affiliation{Department of Physics, Lancaster University,
Lancaster, LA1 4YB, United Kingdom}

\date{\today}

\begin{abstract}
Weyl's law approximates the number of states in a quantum system
by partitioning the energetically accessible phase-space volume
into Planck cells. Here we show that typical resonances in generic
open quantum systems follow a modified, fractal Weyl law, even
though their classical dynamics is not globally chaotic but also
contains domains of regular motion. Besides the obvious
ramifications for quantum decay, this delivers detailed insight
into quantum-to-classical correspondence, a phenomenon which is
poorly understood for generic quantum-dynamical systems.
\end{abstract}
% 05.45.Mt Quantum chaos; semiclassical methods
% 03.65.Sq Semiclassical theories and applications

\maketitle

Phase-space rules provide powerful universal relations for
classical and quantum systems. A time-honored  example is Sabine's
law, originally formulated in the context of room acoustics, which
can be cast into the relation  $\tau_{\rm dwell} =4 V/vA$ for the
mean dwell time  of a classical particle escaping from a
container, expressed in terms of the volume $V$ of the container,
the area $A$ of the opening, and the particle's velocity $v$
\cite{sabine}. In quantum mechanics, Weyl's law approximates the
number ${\cal N}(E)$ of states with energy $E_n<E$  by the number
of Planck cells $h^d$ which fit into the accessible phase-space
volume of the corresponding classical system (here $h$ is Planck's
constant and $d$ is the number of dimensions)  \cite{weyllaw}. For
a quantum particle in a container (e.g., an electron in a quantum
dot), ${\cal N}(E)\propto E^{d/2}$ therefore follows a power law
with a strictly determined exponent.

In reality, quantum systems are open, and the eigenstates acquire
a finite lifetime---the resulting resonance states constitute a
fundamental concept across many fields of physics. However, there
are reasons to believe that the synthesis of both mentioned
phase-space rules in open quantum systems can modify Weyl's law.
Evidence in this direction is provided by systems with globally
chaotic classical dynamics, which exhibit a \emph{fractal} Weyl
law ${\cal N}(E,\tau)\propto c_\tau E^{d_H/2}$ for the number of
resonances with $E_n<E$  and lifetime $\tau_n>\tau$ (where the
cut-off value $\tau$ only enters the shape function $c_\tau$)
\cite{lu:zworski,schomerus,nonnenmacher,keating}. Remarkably, the
number $d_H$ in the exponent is not an integer; instead, it is
given by the dimension of the strange repeller, which for a
chaotic system is a fractal \cite{ott}. While this observation
already proofed suited to instigate a paradigmatic shift of the
study of resonances in open quantum-\emph{chaotic} systems, its
direct practical consequences are necessarily limited: typical
dynamical systems are not globally chaotic, which has profound
consequences on their quantum dynamics \cite{bohigas}.

Here, we show that fractal Weyl laws indeed apply much more
generally to \emph{generic} dynamical systems, for which regular
and chaotic motion coexists in a mixed phase space \cite{ott}.
 The key which reveals the fractal Weyl law is to restrict the
resonance counting to a \emph{window of typical lifetimes}
($\tau<\tau_n<\tau'$, where $\tau\ll\tau_{\rm dwell}\ll\tau'$). We
arrive at this conclusion by a combination of semiclassical
arguments (based on a tailor-made phase-space representation of
resonance wave functions which avoids problems with their mutual
non-orthogonality) with numerical results for a paradigmatic
quantum-dynamical model system, the open kicked rotator
\cite{schomerus,kickedrotator,borgonovi,jacquod}.

\begin{figure}[t]
\includegraphics[width=.48\columnwidth]{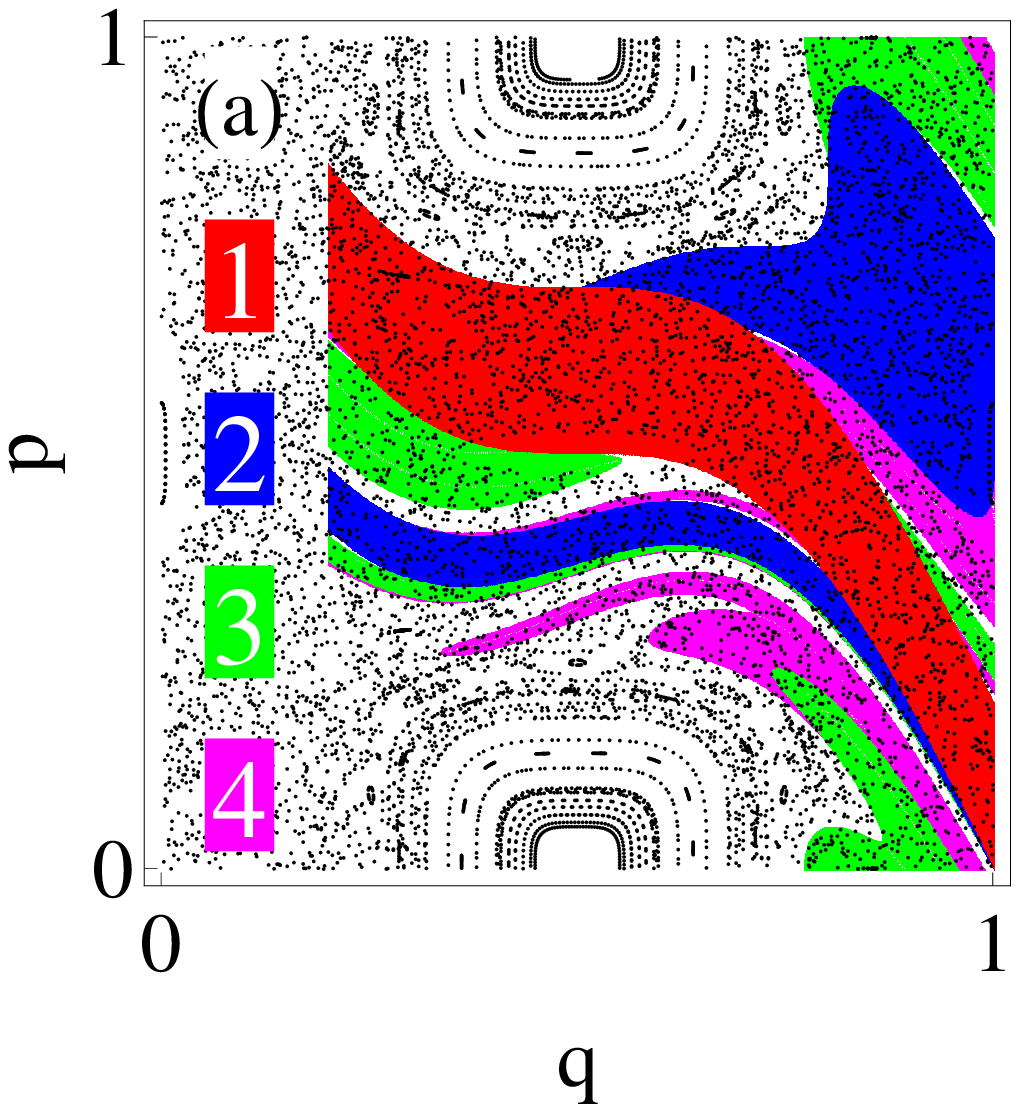}
\includegraphics[width=.48\columnwidth]{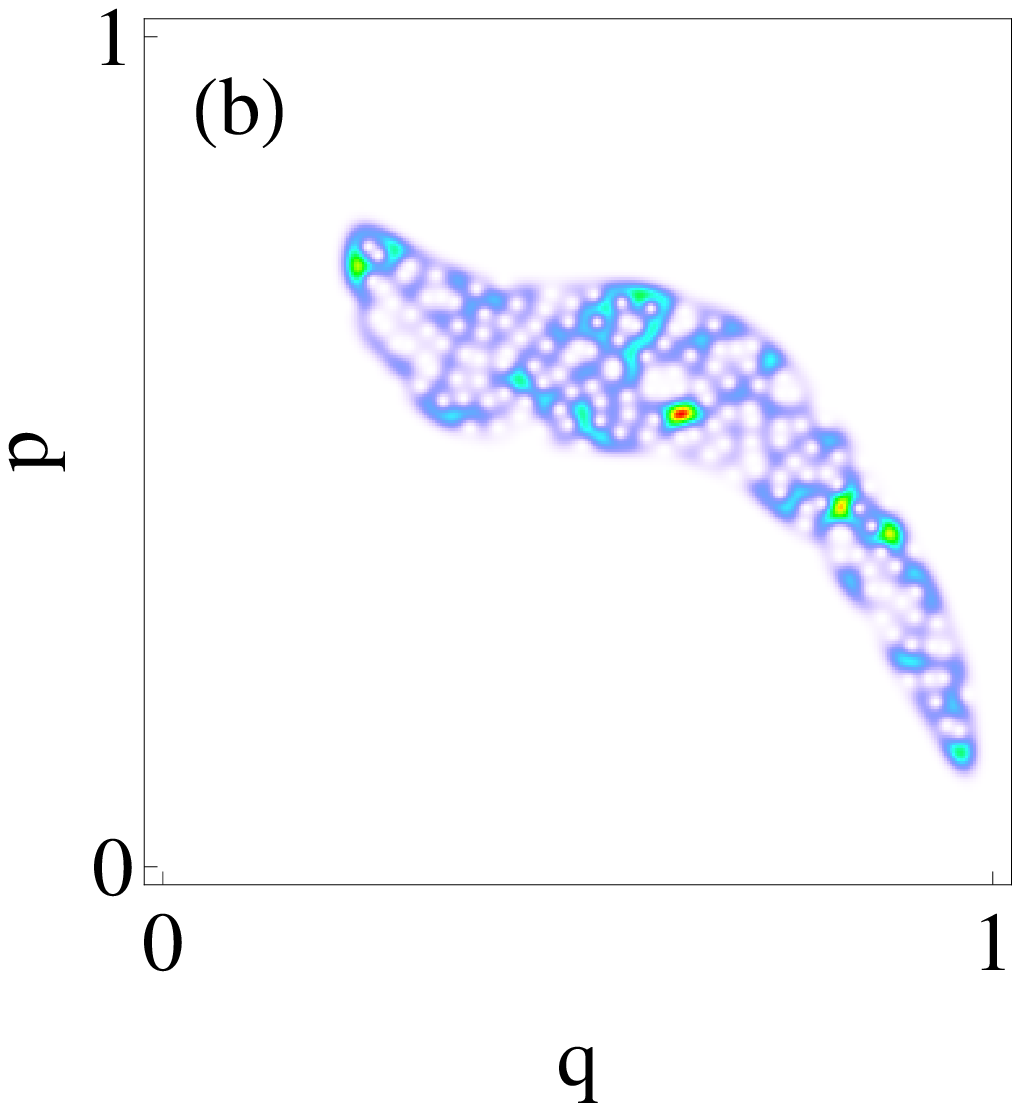}
\\
\includegraphics[width=.48\columnwidth]{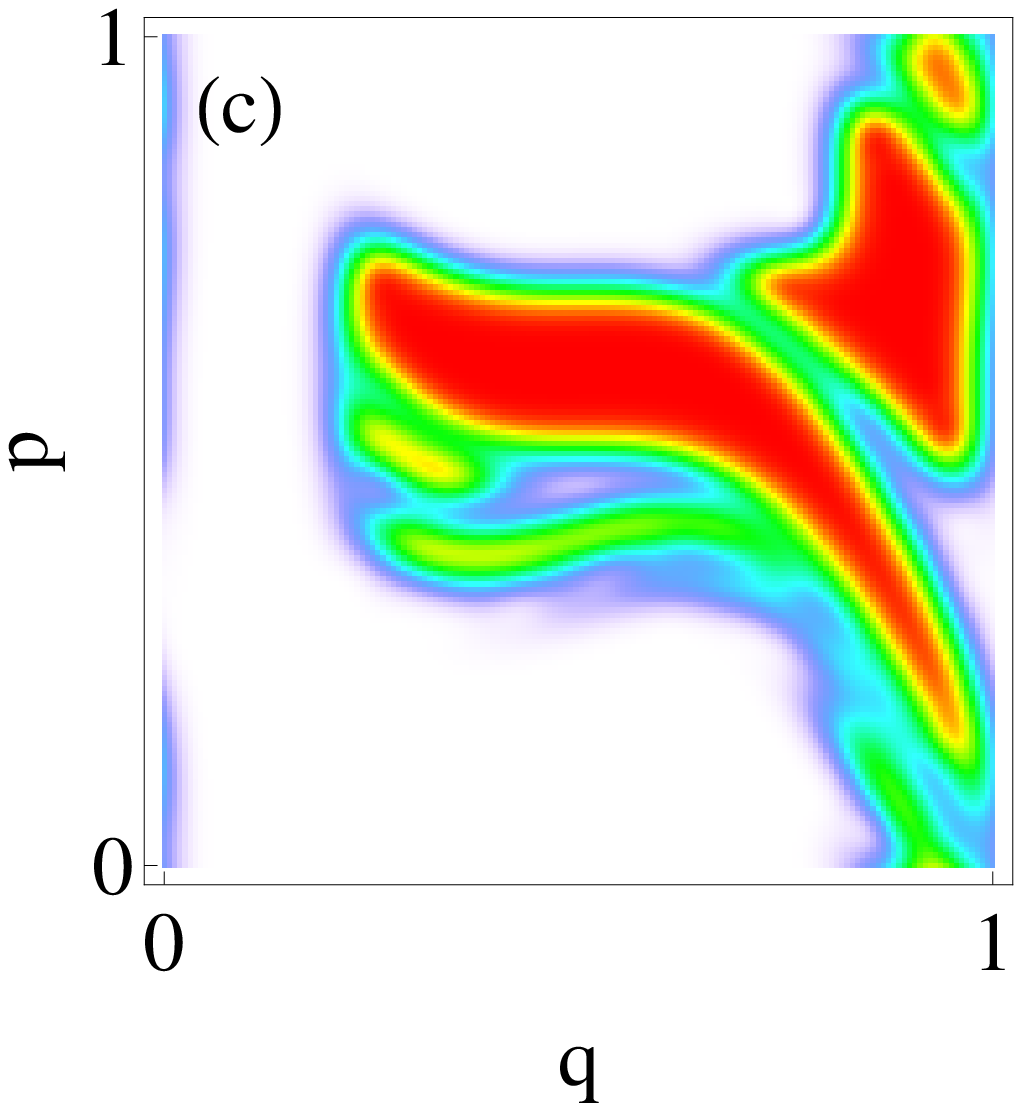}
\includegraphics[width=.48\columnwidth]{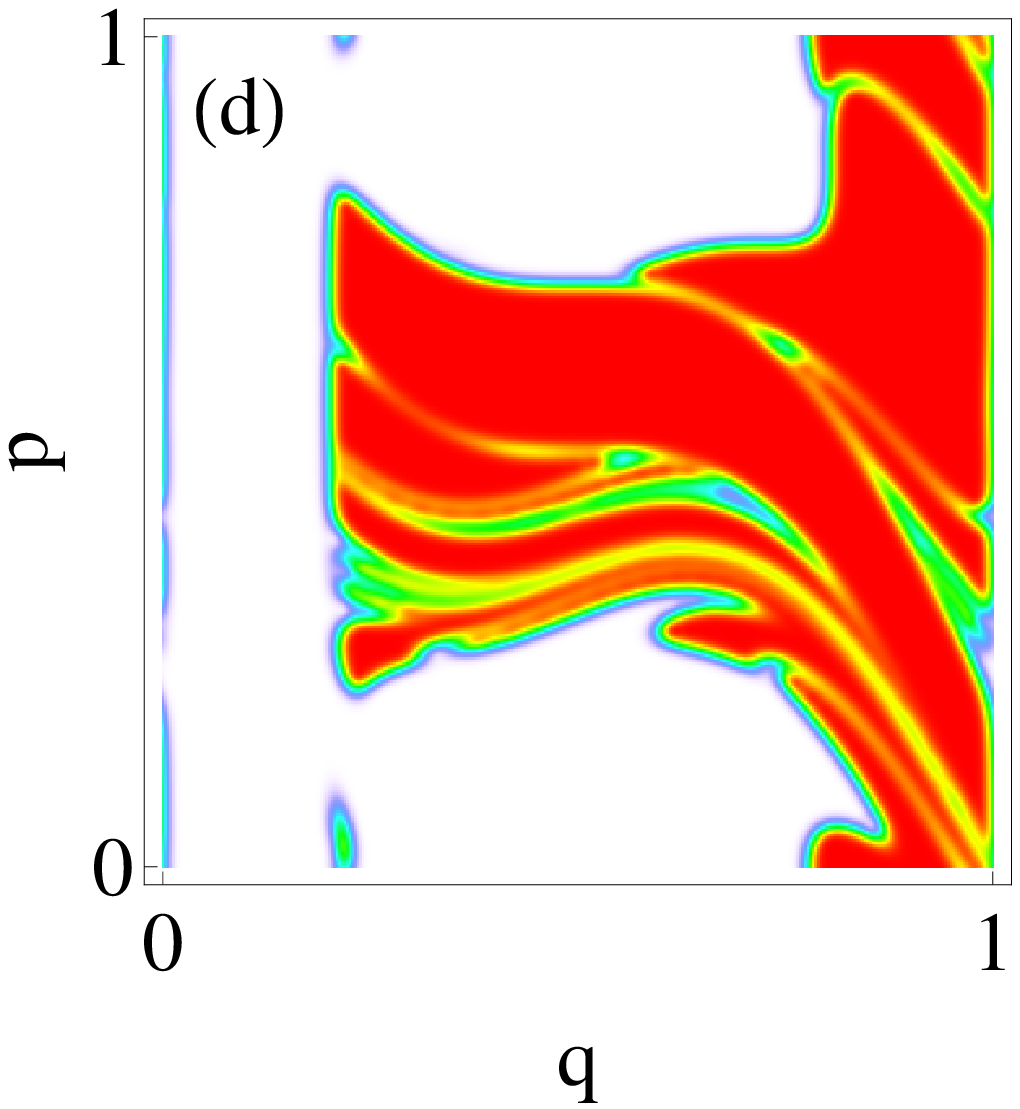}
\\
\includegraphics[width=.48\columnwidth]{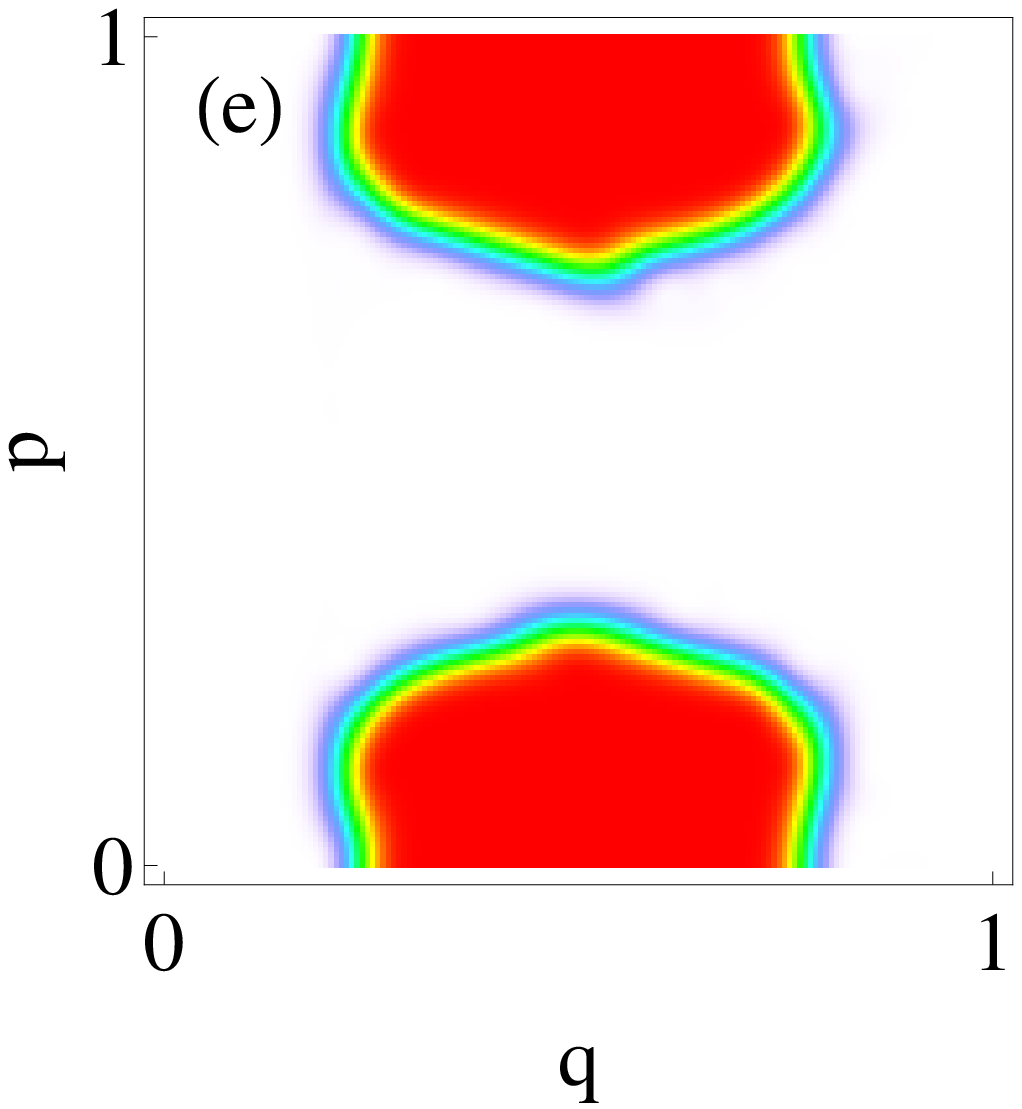}
\includegraphics[width=.48\columnwidth]{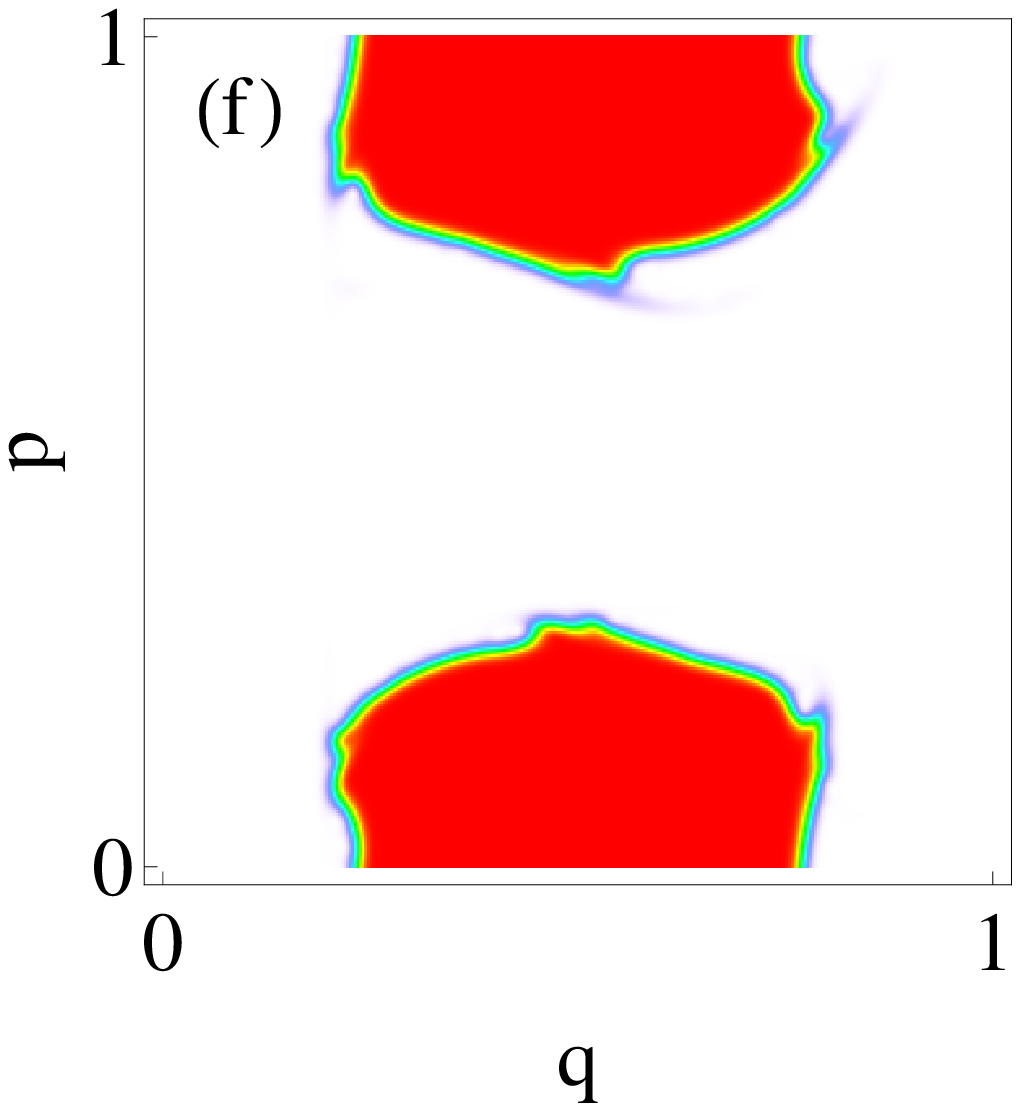}
\caption{(Color) Quantum versus classical escape in a kicked
rotator (kicking strength $k=2.0$) with opening at $0<q<0.2$. (a)
Classical escape zones in phase space $(q,p)$, color coded
according to escape after 1 (red), 2 (blue), 3 (green), or 4
(magenta) iterations. The black dots are trajectory segments in
the closed system, whose phase space is a mixture of regular and
chaotic motion. (b) Husimi representation of short-lived quantum
resonances with decay time $\tau_n< 1/(2\ln 10)$ (decay factor
$|\mu_n|<0.1$), for Hilbert-space dimension $M=1280$. The other
panels show the 'fast' (panels c and d, $|\mu_n|<0.1$)  and 'slow'
(panels e and f,  $|\mu_n|>0.98$) Husimi-Schur representation,
introduced in this work, for $M=160$ (panels c and e) and $M=1280$
(panels d and f).} \label{fig1}
\end{figure}

We start our considerations with some general observations about
quantum-to-classical correspondence. A basic ingredient in the
derivation of the ordinary Weyl law  in closed systems is the
mutual orthogonality of energy eigenstates, which is guaranteed by
the hermiticity of the Hamiltonian (equivalently, the unitarity of
the time-evolution operator). Quantum-to-classical correspondence
can then be exploited, e.g.,  by using a basis of semiclassically
localized states $|x\rangle$ which occupy Planck cells in phase
space $x=(\textbf{q},\textbf{p})$. Resonance wave functions,
however,  overlap with each other because open systems are
necessarily represented by non-normal (neither hermitian nor
unitary) operators. In open quantum maps, for example, resonances
are associated to the spectrum of a truncated unitary matrix
${\cal M}={\cal Q} F {\cal Q}$, composed of the unitary
time-evolution operator $F$ of the closed system and the projector
${\cal Q}={\cal Q}^2$ onto the non-leaky part of Hilbert space
(see Fig.\ \ref{fig1} for an example of an open quantum map).
Because of the truncation, the eigenvalues
$\mu_n=\exp(-iE_n-\gamma_n/2)$ of ${\cal M}$ lie inside the unit
disk of the complex plane, ensuring that the decay rates
$\gamma_n=1/\tau_n$ are positive \cite{remarkonP}. But since
${\cal M}$ is not normal, the associated eigenstates---which now
describe the resonance wave functions---are not orthogonal to each
other. This circumstance, which is the root for the scarceness of
analytical tools in non-normal problems, complicates the task of
exploiting quantum-to-classical correspondence for the purpose of
resonance counting.

We circumvent this problem by applying standard phase space
methods to an alternative spectral decomposition, the Schur
decomposition \cite{schurdecomp} ${\cal M}=UTU^\dagger$, which
delivers the same eigenvalues (occupying the diagonal of the upper
triangular matrix $T$) but associates to them an orthogonal basis
set $U$. For definiteness assume that all eigenvalues are ordered
by their modulus,
$|\mu_1|\leq|\mu_2|\leq|\mu_3|\leq\ldots\leq|\mu_M|$. The first
$r$ rows $u_r$ of $U$ then form a complete basis for the $r$
fastest decaying resonance states. Quantum-to-classical
correspondence can now be exploited by considering the Husimi
representation \cite{husimi} ${\cal H}_r(x)= \sum_{m=1}^r|\langle
x|u_m\rangle|^2$ of this subspace. This  provides insight into the
regions in classical phase space which support the quickly
decaying quantum resonances. An analogous construction can be
based on the opposite ordering,
$|\mu_1|\geq|\mu_2|\geq|\mu_3|\geq\ldots\geq|\mu_M|$, which
focusses on the slowest decaying resonance wave functions. Because
of non-orthogonality, the resulting 'fast' and 'slow' Husimi-Schur
representations carry independent information on the resonance
wave functions.

Figure \ref{fig1} demonstrates the viability of this method for a
paradigm of quantum chaos, the kicked rotator \cite{kickedrotator}
with unitary time-evolution operator
\begin{equation}
F_{nm}=(iM)^{-1/2}
e^{\frac{i\pi}{M}(m-n)^2-\frac{iMk}{4\pi}(\cos\frac{2\pi
n}{M}+\cos\frac{2\pi m}{M})},
 \end{equation}
which classically reduces to the (symmetrized) standard map $p'=
p+k \cos (2 \pi q+\pi p) \mbox{ mod } 1, q'=q+p/2+p'/2  \mbox{ mod
} 1$. The matrix dimension $M=h^{-1}$ determines the inverse
effective Planck constant \cite{remarkonM}, while the kicking
strength $k$ determines the nonlinearity. For $k=0$ the classical
system is integrable, while for $k\gtrsim 7$ the  dynamics is
globally chaotic (resonances in the chaotic variant have been
studied in Refs.\ \cite{borgonovi,schomerus}).

In Figure \ref{fig1}, the kicking strength is set to $k=2$, for
which the phase space of the closed system is mixed, as shown by
trajectory segments (black dots) in panel (a). The color-coded
areas superimposed on the phase space indicate the classical
initial conditions for escape after 1, 2, 3, or 4 iterations when
an opening is placed at $0<q<0.2$ ($\tau_{\rm dwell} = 5$). Panel
(b) shows the ordinary Husimi representation of quickly decaying
resonance eigenfunctions with $0<|\mu_n|<0.1$ ($M=1280$). Panels
(c) and (d) show the 'fast' Husimi-Schur representation
($0<|\mu_n|<0.1$, $M=160$ and $M=1280$), while panels (e) and (f)
display the `slow' Husimi-Schur representation ($0.98<|\mu_n|<1$;
$M=160$ and $M=1280$) \cite{remarkonevals}.

The ordinary Husimi representation demonstrates that the quickly
decaying resonance eigenfunctions are all localized in the region
of escape after a single iteration. However, since these
eigenfunctions strongly overlap, Planck-cell partitioning their
phase-space support significantly underestimates their number
(predicting $\simeq 25$ instead of $r=44$ short-lived states for
$M=160$, and $\simeq 200$ instead of $r=568$ short-lived states
for $M=1280$). In contrast, the fast Husimi-Schur representation
clearly maps out the classical escape zones and uncovers that the
domain of quantum-to-classical correspondence increases with
increasing $M$. The slow Husimi-Schur representation maps out
stable phase-space regions that are classically decoupled from the
opening by impenetrable dynamical barriers; these regions do not
significantly change as $M$ increases.

By construction, Planck-cell partitioning of the support of the Husimi-Schur
representations accurately estimates the underlying number of resonances. The
fast Husimi representation therefore uncovers a  proliferation of anomalously
short-lived states driven by the emerging quantum-to-classical correspondence
--- among the total count of $M$ resonances, the fraction of short-lived
states increases as $M$ increases. On the other hand, the slow
Husimi representation shows that the fraction of anomalously
long-living resonances supported by classically uncoupled regions
remains fixed, which corresponds to the ordinary Weyl law for this
effectively closed-off part of the system. Crucially, we are led
to conclude that the remaining fraction of resonances with
\emph{typical} lifetime (chosen here to satisfy
$0.1<|\mu_n|<0.98$) {\em decreases} as $M$ increases, and
therefore cannot follow an ordinary Weyl law.

A detailed understanding of the resonance distribution can be
obtained by counting the resonances in fixed lifetime windows and
comparing these counts for different values of $M$. We adopt
probabilistic terminology and proceed in two steps. In the first
step we determine the fraction $P(\mu)={\rm prob}(|\mu_n|>\mu)$ of
resonances with lifetime exceeding a lower threshold
$\tau=-1/(2\ln \mu)$. This defines a monotonously decreasing
function interpolating between $P(0)=1$ and $P(1)=0$. In the
second step, we extract the fraction of resonances within an
interval of typical lifetimes ($0.1<|\mu_n|<0.98$), which follows
from $P_{\rm typ}={\rm prob}(|\mu_n|\in
[0.1,0.98])=P(0.1)-P(0.98)$ (our results and conclusions do not
depend on the chosen window as long as it stays in the range of
typical life times).

\begin{figure}[t]
\includegraphics[width=\columnwidth]{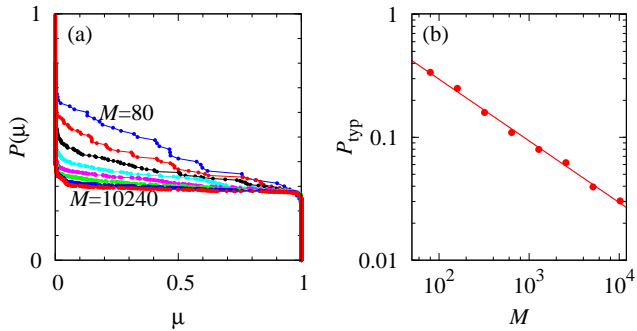}
\caption{(Color online) Fractal Weyl law for resonances in the
open kicked rotator with a mixed classical phase space (kicking
strength $k=2.0$, opening  at $0<q<0.2$). (a): Fraction $P(\mu)$
of resonances with decay factor $|\mu_n|>\mu$, corresponding to a
lifetime $\tau_n> -1/(2\ln\mu)$. (b) Fraction $P_{\rm typ}$ of
resonances in a band of typical lifetimes ($0.1<|\mu_n|<0.98$), as
a function of dimensionless system size $M$. In this
double-logarithmic representation, the fractal Weyl law results in
a linear dependence with a negative slope.
 } \label{fig2}
\end{figure}

Numerical results for the system with opening at $0<q<0.2$ are
shown in Fig.\ \ref{fig2}. Panel (a) shows $P(\mu)$ for various
values of $M$. As a function of $\mu$, $P$ decreases very sharply
at the two extreme ends of the graph. At $\mu \simeq 0$ we witness
the influence of extremely short-lived resonances, while at $\mu
\simeq 1$ we observe the states which have very long lifetimes.
Applicability of the ordinary Weyl law would entail that modulo
small fluctuations, $P(\mu)$ is independent of $M$, since the
uncertainty-limited resolution of phase space increases uniformly
when the Planck cell shrinks. The plot, however, shows that the
body of the function $P$ drops as $M$ increases. This is due to
the proliferation of the short-lived resonances ($\mu\simeq 0$),
whose relative fraction among all resonances increases with
increasing $M$, in agreement with the expanding domain of support
of the fast Husimi-Schur representation. In the region of
long-living states $(\mu \simeq 1)$, on the other hand, $P$ does
not depend significantly on $M$, in agreement with the observed
$M$-independent support of long-living resonances in the slow
Husimi-Schur representation.

Complementing these trends for short and long lifetimes, the body
of $P$ becomes flatter as $M$ increases. As shown in Fig.\
\ref{fig2}(b), the fraction $P_{\rm typ}$ of resonance with
typical lifetime therefore decreases  with increasing $M$. The
linear fit in this  double-logarithmic plot demonstrates that this
trend closely follows a power law $P_{\rm typ}\propto M^{-0.50}$.
The typical resonances therefore  obey a fractal Weyl law, ${\cal
N}_{\rm typ}=M P_{\rm typ} \propto M^{0.50}$.

\begin{figure}[t]
\includegraphics[width=\columnwidth]{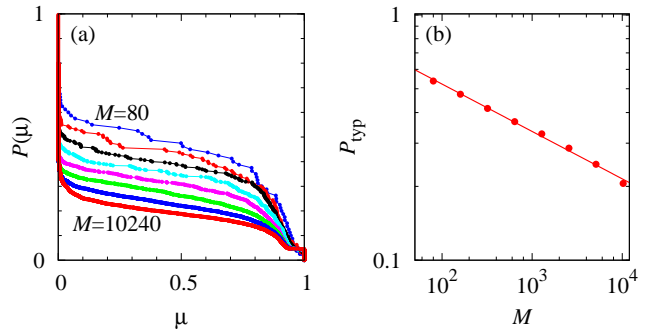}
\caption{(Color online) Same as Fig.\ \ref{fig2}, but with opening
shifted to $0.2<q<0.4$.
 } \label{fig3}
\end{figure}

\begin{figure}[b]
\includegraphics[width=.9\columnwidth]{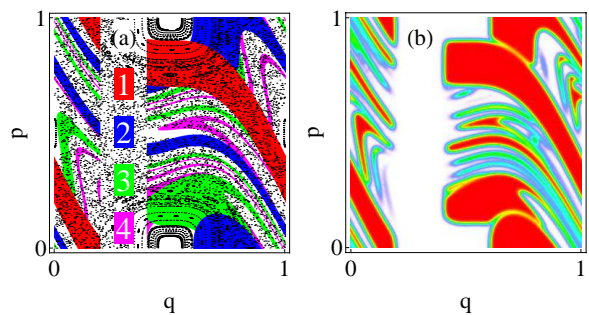}
\caption{(Color) (a) Classical escape zones and (b) fast
Husimi-Schur representation of short-lived quantum resonances
($M=1280$) for the kicked rotator of Fig.\ \ref{fig1},  with
opening shifted to $0.2<q<0.4$. } \label{fig4}
\end{figure}

Figure \ref{fig3} shows that the fractal Weyl law remains intact
when the opening is shifted to $0.2<q<0.4$, so that it couples to
a larger part of the regular regions in phase space [see Fig.\
\ref{fig4}(a)]. In this case $P_{\rm typ}\propto M^{-0.19}$, and
therefore ${\cal N}_{\rm typ}\propto M^{0.81}$. Compared to the
situation in Fig.\ \ref{fig2}, the fraction of long-living states
is now reduced, in keeping with the shrunken size of the
classically uncoupled phase-space region. The fast Husimi-Schur
representation [Fig.\ \ref{fig4}(b)] shows that the newly coupled
parts support additional  short-lived resonances. Consequently,
the states with typical lifetime are still associated to the
chaotic regions.

For globally chaotic systems, the fractal Weyl law can be
understood by associating the proliferation of short-lived
resonance to quasi-deterministic decay following classical escape
routes \cite{schomerus}. This quasi-deterministic decay requires
classical-to-quantum correspondence, which is lost exponentially
on a time scale given by the Ehrenfest time
$\tau_E=\lambda^{-1}\ln M$. The probability to reside within the
system decays exponentially, too, and is governed by $\tau_{\rm
dwell}$. The power law for the fractal Weyl law therefore arises
from the combination of two exponential laws, based on the
relation $\exp(-\tau_E/\tau_{\rm dwell})=M^{-1/\lambda\tau_{\rm
dwell}}$ for the part of phase space where classical-to-quantum
correspondence does not apply; this region supports ${\cal
N}\propto M^{d-1/\lambda\tau_{\rm dwell}}$ resonances (delivering
an accurate estimate for $d_H$ if the opening is sufficiently
small \cite{ott}). For generic dynamical systems, our results
confirm the association of short-lived resonances to
quasi-deterministic escape routes, while the resonances of typical
lifetime are now associated to chaotic regions  dominated by
sticking motion, where classical power-law decay $\propto
t^{-\alpha}$ rule in place of the exponential decay
\cite{sticking}. This can be reconciled with the fractal Weyl law
when one assumes that the loss of quantum-to-classical
correspondence in these regions is similarly modified into a power
law $\propto t^\beta$, so that the Ehrenfest time takes the
algebraic form $\tau_E\propto M^{1/\beta}$ \cite{algebraic}. The
fractal Weyl law then arises from the combination of two power
laws, based on the relation $\tau_E^{-\alpha}\propto
M^{-\alpha/\beta}$ for the part of phase space where
classical-to-quantum correspondence does not apply and dynamics is
dominated by transient sticking motion.

In summary, we have shown that typical resonances  in  open
quantum systems obey a modified, fractal Weyl law, a phenomenon
previously associated only to the non-generic case of systems with
a globally chaotic classical limit. We unravelled this law by
introducing the concept of a Husimi-Schur representation, a
phase-space representation of resonance wave functions which
captures maximal information on quantum-to-classical
correspondence (circumventing the problem that resonance
eigenfunctions are not orthogonal to each other). The formation of
anomalously short-lived states that drives the departure from the
ordinary Weyl law originates in quasi-deterministic decay along
classical escape routes, whose phase-space support expands as one
approaches the classical limit. The fractal Weyl law emerges from
the interplay of the transient sticking of chaotic trajectories to
the stable components and the algebraic violation of
quantum-to-classical correspondence, two mechanisms which are
intimately related to a mixed phase space.

The kicked rotator used here for illustration can be realized with
atoms that are driven by pulsed optical waves \cite{atomoptics}.
Experimentally, direct evidence of the fractal Weyl law is more
likely to come from autonomous (non-driven) systems such as
microwave resonators, in which advanced techniques allow for  the
accurate determination of complex resonance frequencies
\cite{KHMS08}; a mixed phase space is obtained for any generic
smooth resonator shape.  While the details of a mixed phase space
constitute a unique fingerprint of a given dynamical systems, the
general features  (a hierarchy of stability islands embedded into
chaotic domains where long-time transients arise from sticking
motion) are remarkably robust. Based on this general phenomenology
we expect that the fractal Weyl law for typical resonances is a
generic feature of open quantum-dynamical systems.

This work was supported by the European Commission via  Marie
Curie Excellence Grant No. MEXT-CT-2005-023778.


\begin{thebibliography}{28}
\bibitem{remarkonP} The projector on the leaky part is ${\cal P}=1-{\cal Q}$, and $\mbox{rank}\,{\cal P} = N$
determines the number of open decay channels. Since ${\cal M}{\cal
P}=0$, there are also $N$ trivial eigenvalues with $\mu_n=0$ and
therefore $\tau_n=0$, corresponding to states that immediately
decay through the opening.
\bibitem{remarkonevals} In  Fig.\ \ref{fig1}(b)-(d) and  Fig.\ \ref{fig4}(b) we omit for clarity the $N$ trivial
eigenvalues with $\mu_n=0$, which uniformly cover the opening.
\end{thebibliography}
\end{document}